\documentclass[ ]{copernicus2}

\usepackage{color}
\usepackage[T1]{fontenc}

\begin{document}

\title{Topside Reconnection
}

{\author[1,3]{R. A. Treumann}
\author[2]{W. Baumjohann$^a$}
\affil[1]{International Space Science Institute, Bern, Switzerland}
\affil[2]{Space Research Institute, Austrian Academy of Sciences, Graz, Austria}
\affil[3]{Geophysics Department, Ludwig-Maximilians-University Munich, Germany\protect\\
Correspondence to: Wolfgang.Baumjohann@oeaw.ac.at}

}

\runningtitle{Topside Reconnection}

\runningauthor{R. A. Treumann \& W. Baumjohann}

\received{ }
\pubdiscuss{ } 
\revised{ }
\accepted{ }
\published{ }


\firstpage{1}

\maketitle

  

\noindent\textbf{Abstract}.-- 
{It is proposed that reconnection would be a main mechanism governing the plasma processes on  auroral time scales in the topside ionosphere/high-latitude magnetosphere transition. It occurs in the downward current region between two narrow parallel closely spaced though separated downward current sheets. The field-aligned currents are carried by upward cold upper-ionospheric electrons closing the upward current in an adjacent region. This local process does primarily not affect the ambient field but generates an anomalous diffusivity.  } 

In a recent paper \citep{treumann2017} we suggested that strong-guide field reconnection may play a role in the generation of radiation in the topside auroral ionosphere by the electron cyclotron maser instability mechanism (ECMI) \citep{wu1979,melrose1985,treumann2006}. This idea was based on the assumption that at the boundary between the upward and downward current regions the magnetic fields of kinetic Alfv\'en waves might undergo reconnection, causing electron exhausts which are similar to electron holes while being of larger scale. Such a mechanism seems to be appropriate to explain intense emission in the auroral kilometric radiation band \citep{gurnett1974}. In a subsequent paper we explored a particular model of stronger amplification of the radiation if electron pairing would occur in the vicinity of the electron mirror points along the strong auroral magnetic field. This brings up the question whether, independent of the generation of radiation, reconnection might not be the dominant process of plasma dynamics in the topside auroral region.

The topside auroral region is characterized by a number of properties which at first glance do not seem in favour of reconnection. In order to have reconnection one needs contact between anti-parallel magnetic fields and plasma inflow perpendicular to the field, as is inherent to all the basic reconnection models \citep[see][for reviews separated by 40 years]{vasyliunas1975,treumann2013}. The auroral region, at the contrary, hosts a very strong magnetic field which on the scales of aurora is parallel (except for some weak inclination and systematic geographic variation). It does not change sign across the auroral region on one hemisphere. Thus it  seems highly improbable that such a field would undergo any reconnection and rearrange in some violent manner. Indeed, it does not. In order to rearrange it requires very strong external basically mechanical forces to twist or wrap it around. Such forces would be related to extremely strong currents which in the topside ionosphere-magnetosphere transition renders them completely improbable. {They require conditions which are presumably realized in the lower solar atmosphere where the solar photospheric convection network rotates the frozen-in magnetic field at frequency of few minutes around causing the field to become highly warped into a spiral which stretches out into the corona and solar wind.} The strong geomagnetic field in topside auroral region is in contrast fixed to the inert ionosphere and body of the earth. Deforming it substantially requires very strong outer forces which happens very rarely. It may, however, play a decisive role as catalyst of reconnection caused by other means. 

The auroral region is comparably narrow in {(geographic/geomagnetic)} latitude while somewhat broader in its longitudinal extension. In each auroral event, it divides into two sections, one carrying \emph{downward} fluxes of medium energy $\epsilon_e\sim 10$ keV electrons corresponding to \emph{upward} field-aligned currents, the other hosting \emph{upward} electron fluxes of energies $\epsilon_e\lesssim$ 1 keV corresponding to \emph{downward} field-aligned currents. There is a spectrum of magnetic variations \citep{labelle2002} partially related to these currents which is interpreted differently in terms of waves below the local electron cyclotron frequency $\omega<\omega_c$ which at the lower spectral end has substantial amplitudes $\delta B$ usually interpreted as caused by fluctuations of the main field $B_0$ imposed by the magnetosphere. Still, the amplitudes are small in the sense that $\delta B^2\ll B_0^2$. The former section has several times larger spatial extension than the latter. The two regions occur always in tandem, {typical for a closed current-return current system} and, in most cases, not as a single current pair but in groups of several up-down pairs, generally being separated by a region of no auroral current activity. The phenomenology and models have been described in \citet[][chpts.1-8]{paschmann2003}. 

\begin{figure*}[t!]
\centerline{\includegraphics[width=0.75\textwidth,clip=]{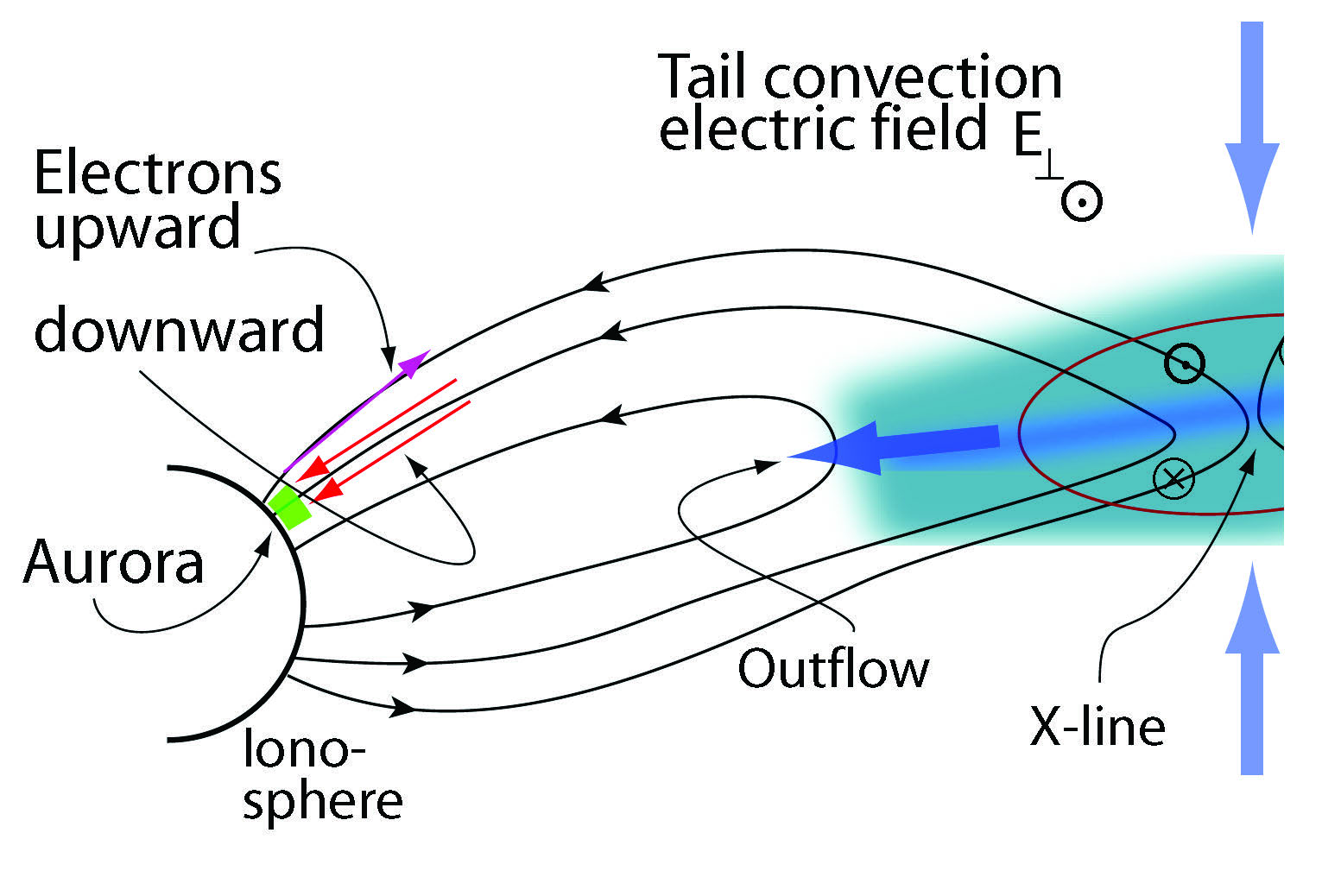}}
\caption{Schematic of connection between tail reconnection and auroral topside current system for one single tail X-point \citep[after][]{treumann2011a}. Downward and upward field-aligned electron fluxes are indicated in the topside ionosphere. They correspond to upward and downward field-aligned currents. Naturally, due to the geometry of the tailward source, the downward fluxes are distributed over a large spatial domain, while the returning upward fluxes occupy a narrow latitudinal interval only on the northern edge.} \label{fig1}
\end{figure*}

There is no obvious {\emph{local} reason for field-aligned currents in the topside auroral ionosphere to be dispersed in the manner observed}. Their most reasonable driving source is reconnection in the tail current sheet, however, which is well established. Upward currents/downward electrons originate from electron acceleration in the central tail plasma sheet. They flow down along the newly reconnected closed magnetic field into auroral latitudes causing the upward field aligned currents. Upward low energy electron fluxes belong to the downward closure currents and result from moderately accelerated ionospheric electrons present here in sufficiently large numbers. How this acceleration happens in detail remains unclear but can be taken as an observational fact. More than one up-down pair of currents indicates multiple tail reconnection.  It seems that this is the only causally satisfactory picture. (Its gross geometry is depicted in Fig. \ref{fig1}.)

Some of the most violent auroral processes result from the dynamics of the downward/upward electron fluxes and the related upward/downward field-aligned currents. (A full sequence of FAST observations when crossing a topside active auroral region during a substorm is given in Fig. \ref{fig2}.)  Of course, since reconnection in the tail is non-stationary, its longer temporal scale folds into the internal processes caused by the fluxes and currents. It modifies those while can be considered as waves flowing along the field with non-stationary currents coming in field-aligned electromagnetic wave pulses. These are assumed as belonging to one of the (kinetic) Alfv\'en modes. Thus, on the time-scale of the latter, auroral dynamics will be related to the electromagnetic stability of the field-aligned current pulses. 

\section{Field-aligned downward current sheets}
The observations concerning auroral electron fluxes and the related currents are the following:

\subsection{Upward current region properties}
Downward electrons/upward currents occupy an extended spatial interval of  low density and barely structured fluxes. The variation of the perpendicular (to the main field) magnetic field is smooth; it changes about linearly from $-\delta B_\perp$ to $+\delta B_\perp$ signalling that the spacecraft has crossed a homogeneous broad structureless upward sheet current carried by the as well structureless medium energy electron flux which, in the energy-time spectrum occupies a narrow band of constant energy and small energy spread.

Absence of an ionospheric electron background at (FAST) spacecraft altitude either suggests that the ionosphere does not reach up to those altitudes ($\sim2000- 3000$ km) which sometimes, in a diffusive model of the ionosphere, is interpreted as presence of a field aligned electric potential which holds the ionospheric electrons down while attracting magnetospheric electrons. The validity of such an assumption can be questioned in terms of tail reconnection as the inflow of reconnection accelerated electrons from the tail does not require such an electric potential field, the origin of which is  difficult to justify over a region of upward current extension while being natural when considering tail reconnection where it simply maps the large reconnection affected interval of the cross tail current down into the ionosphere. The small number of downward electrons does not require any presence of electric fields. The flux consists of nearly mono-energetic auroral electrons. These form a field-aligned beam and are accompanied by observed low frequency Langmuir-wave excitation which allows for the determination of the beam density being roughly $N_\downarrow\approx 10^6$ m$^{-3}$ (one electron per cubic centimetre). 

\subsection{Upward topside electrons}
Figure \ref{fig3} shows simultaneous upward/downward FAST measurements of electron fluxes when crossing a very active substorm topside auroral ionosphere. The upward-electron downward-current region behaves  differently. Its spatial extension is narrow. In view of the electron flux it consists of a large number of very closely spaced spikes. The flux in each spike (generally) maximizes at the lowest energies $\epsilon_e\lesssim 0.1$ keV.  Electron number densities are high estimated to be around $N_\uparrow \sim 10^7$ m$^{-3}$ (ten per cubic centimetre) or higher. The total integrated up and down currents must be similar for perfect closure. This is however not guaranteed for the divergence of currents in the ionosphere perpendicular to the magnetic field, current dissipation, and the high spatial structuring of the downward currents such that it cannot be checked whether the indication of the different downward current sheets all belong to closure of the single upward current. Some uncertainty in comparison remains, which however for our purposes does not matter.

The important observation is the high local structure of the downward currents, their obvious spatial closeness, and their differences in energies and flux level which is reflected in both the flux fluctuations across the narrow downward current region, and in the high spatial fluctuation of the main-field-perpendicular magnetic component $\mathbf{b}_\perp$ {(from here on denoting the magnetic variation $\delta B_\perp= b_\perp$)} which indicates the crossing of many downward current sheets or filaments.  All these downward currents flow parallel while being closely spaced in the direction perpendicular to the main field $\mathbf{B}_0$.  Electrodynamics requires that they should attract each other and merge. Why is this not happening in the auroral downward current region?

One might argue that the acceleration of electrons in the ionosphere below observation altitude is probably highly localized, depending on processes in the resistive ionospheric plasma.  Therefore there would be no need for upward escaping electrons to merge laterally. This argument is invalid because they transport current. Lorentz attraction forces the currents to approach each other to form a broad unstructured downward current sheet. This is, however, inhibited by the strong main auroral geomagnetic field $B_0$. The argument that this should also happen in the upward current region fails because the current sheet there is broad by its origin from the tail reconnection site.
\begin{figure*}[t!]
\centerline{\includegraphics[width=0.75\textwidth,clip=]{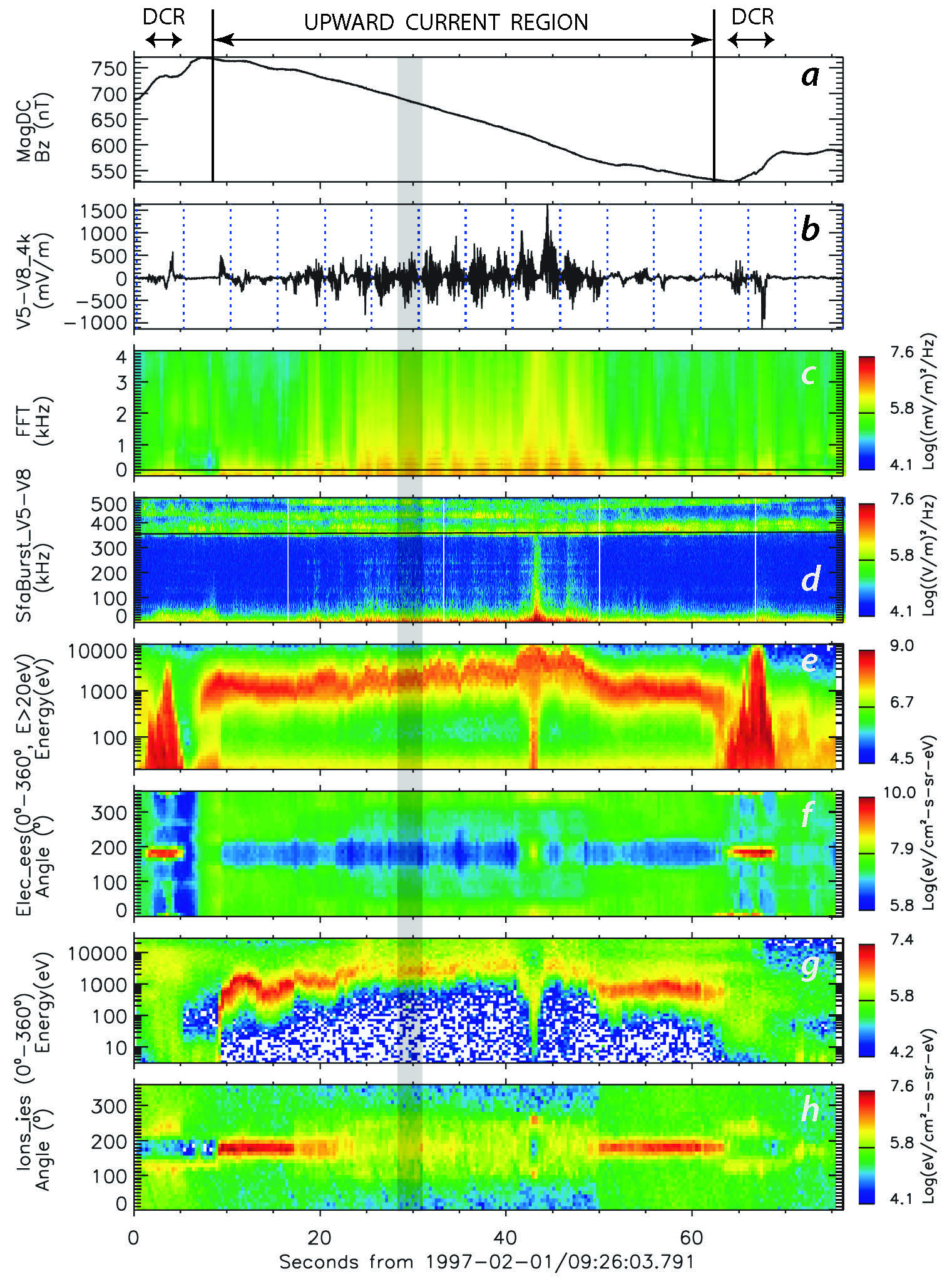}}
\caption{Full sequence of FAST measurements across dow-up-auroral current system on 02-01-1997  \citep[after][]{treumann2011}. {$(a)$ Magnetic field component $b_\perp$ transverse to main field $B_0$, $(b)$ electric field fluctuation wave form $\delta E$, $(c)$ low frequency electric fluctuation spectrum, $(d)$ high frequency electric spectrum, showing emission of auroral kilometric radiation bands $(e)$ electron energy spectrum, $(f)$ electron flux versus pitch-angle, $(g)$ ion energy flux, $(h)$ ion flux versus pitch-angle. The most intriguing part here is the smoothness of the magnetic signature of the upward current in its linear course showing that the upward current is a broad homogeneous current sheet. The downward current region (DCR) flanks the upward current region to both its sides, is comparably narrow in its spatial extent, and exhibits strong current and flux variations. This is seen in the electron flux panels $(e-f)$. Downward fluxes around few keV are relatively smooth indicating a relatively stable tail reconnection over observation time, upward fluxes have maximum at low energy and are highly variable in time and space.}The magnetic field being the integrated response to the spatial flux fluctuations exhibits a much smoother course which is inverse with respect to that of the upward current thus indicating the reversed current direction. Note the low energies of the upward electron fluxes in panel five as well as the clear separation of upward and downward fluxes as seen in the left part.} \label{fig2}
\end{figure*}

Since anti-parallel currents reject each other the transition region between upward and downward currents is quiet. {This is seen in panels $(b - f)$ of Fig. \ref{fig2} and} is in contrast to our previous investigation where we assumed that reconnection would happen there between parallel kinetic Alfv\'en waves. The Lorentz force between two equally strong sheet currents $\mathbf{J}_\|$ is 
\begin{equation}
\mathbf{J}_\| \times \mathbf{b}_\perp= -\nabla_\perp b_\perp^{2}/\mu_0  
\end{equation}
where $\mathbf{b}_\perp$ is the magnetic field between the two currents {(in the following we suppress the index $\perp$ on the magnetic field component $\mathbf{b}$)}, and  $\nabla_\perp$ refers to the gradient in the direction from current sheet to current sheet. The current consists, however, of gyrating electrons whose Lorentz force is the cross product of the azimuthal gyration speed times the very strong stationary field with gradient $\nabla_c$ taken only over the gyro-radius $r_{ce}=v_{e\perp}/\omega_{ce}$ of the electrons. For a separation of the sheet current exceeding the electron gyroradius and low current density the sheet currents will approach each other only on very long diffusive time scales of no interest. For a thin current sheet only a few gyroradii thick the condition for this time to be long is simply that the electron inertial length exceeds the gyroradius or
\begin{equation}
v_{e\perp}/c\ll\omega_{ce}/\omega_e
\end{equation}
which holds under very weak conditions in the topside auroral ionosphere. This implies that downward current sheets separated by say an electron inertial length $\lambda_e=c/\omega_e$ will not merge under no circumstance. They remain separated over the observational spacecraft crossing time scales. It is their secondary magnetic field $\mathbf{b}$ which will undergo reconnection without affecting the ambient magnetic field which just serves as guide field directed along the current flow. This distinguishes topside reconnection from other guide field mediated reconnection. {One may note, however, that sometimes in simulations when plasmoids form \citep[cf., e.g.,][and others]{malara1991} parallel currents apparently do not attract each other. This happens, when the Lorentz force between the parallel currents does not overcome the mechanical forces exerted by the massive plasmoids, i.e. forces induces by their inertia and impulse. The Lorentz force is then too weak to push the parallel currents toward each other, an effect which can also be observed in highly turbulent plasmas. Such cases, when the currents remain close enough will, by the mechanism proposed below, be subject to reconnection between the opposing fields of the parallel current, leading to a cascade in the current structure towards smaller scales and to local reconnection as a main dissipation process of magnetic and turbulent energy.}

\subsection{Kinetic (shear) Alfv\'en waves}
In the complementary wave picture of field-aligned currents in the auroral region, the current is carried by (kinetic) Alfv\'en waves in the frequency range well below the local ion-cyclotron frequency. In addition, a large number of low-frequency electromagnetic waves are known to be present there \citep{labelle2002}. We are in the downward current region with highly sheared upward particle flow along the magnetic field consisting of moderately fast electrons and much slower ions. Such flows are capable of generating Alfv\'en waves \citep{hasegawa1982} on perpendicular scales of the ion inertial length $\lambda_i=c/\omega_i=\lambda_e\sqrt{m_i/m_e}$ and below and long wavelength parallel to the ambient field. For the current-carrying electrons such waves are about stationary magnetic structures. 

These Alfv\'en waves cannot be body waves like in the solar wind \citep{goldstein2005,narita2020} because they are strictly limited to the narrow field-aligned current sheets. Since they propagate in the strong auroral geomagnetic field, they are rather different from the usual kind of kinetic Alfv\'en waves which one refers to in solar wind turbulence \citep{goldstein2005}, where the magnetic field ist very weak and the turbulence is dominated by the mechanics of the flow \citep{maiorano2020}. There the ion-temperature plays an important role imposing kinetic effects on the wave. 

Under auroral conditions, in particular close to the ionosphere, the magnetic field is so strong that thermal ion effects on the wave are barely important. Their mass effect enters the Alfv\'en speed. Instead, however, under those conditions electron inertia on scales $\lambda_i\sim\Delta\gtrsim\lambda_e= c/\omega_e$ below the ion scale comes into play. For sufficiently narrow field-aligned current sheets of width the order of inertial scales, the field does not allow the electrons to leave their flux tube unless they have large perpendicular moment. Field-aligned electrons remain inside their gyration flux tube, and the currents cannot react to merge with neighbouring parallel current sheets. The Lorentz force on the field-aligned current in the magnetic field of its neighbour is not strong enough to move the currents. In this case the kinetic Alfv\'en waves transporting the currents in pulses become inertially dominated with dispersion relation
\begin{equation}
\omega^2= \frac{k_\|^2V_A^2\big(1+k_\perp^2\rho_i^2\big)}{1+k_\perp^2\lambda_e^2}
\end{equation}
where $\rho_i$ is some modified ion gyro-radius \citep[cf., e.g.][]{baumjohann1996} containing kinetic temperature contributions. For the cold ions in the topside auroral ionosphere the term containing $\rho_i$ in the numerator vanishes. The kinetic Alfv\'en wave under those conditions becomes an inertial or shear wave. It propagates at a reduced though still fast speed along the magnetic field, being of very long parallel wavelength. It also propagates slowly perpendicular to the magnetic field at short wavelength $\lambda_\perp\sim\lambda_e/2\pi$. It is, in principle, this wave which carries the current. Thus the current is not stationary on time scales long compared to the inverse frequency $\Delta t>\omega^{-1}$ but can be considered stationary for shorter time scales  $\omega\Delta t < 1$. 

The above dispersion relation, neglecting the ion contribution in the numerator, gives the well known relations for the parallel and perpendicular energy transport in the shear wave
\begin{equation}
\frac{\partial\omega}{\partial k_\|}=\frac{V_A}{\sqrt{1+k_\perp^2\lambda_e^2}},\qquad \frac{\partial\omega}{\partial k_\perp}=-\frac{\partial\omega}{\partial k_\|}\frac{k_\|}{k_\perp} \frac{1}{\big[1+1/(k_\perp\lambda_e)^{2}\big]}
\end{equation}
Energy transport in the perpendicular direction is smaller than parallel by the ratio of wave numbers.

Repeating that we are in the downward current upward electron flux region causality requires that the upward electrons carry information from the ionosphere to the magnetosphere. Hence the kinetic Alfv\'en waves in this region also propagate upward being produced in the topside by the transverse shear on ion-inertial scales below $\lambda_i$. 

\begin{figure*}[t!]
\centerline{\includegraphics[width=0.8\textwidth,clip=]{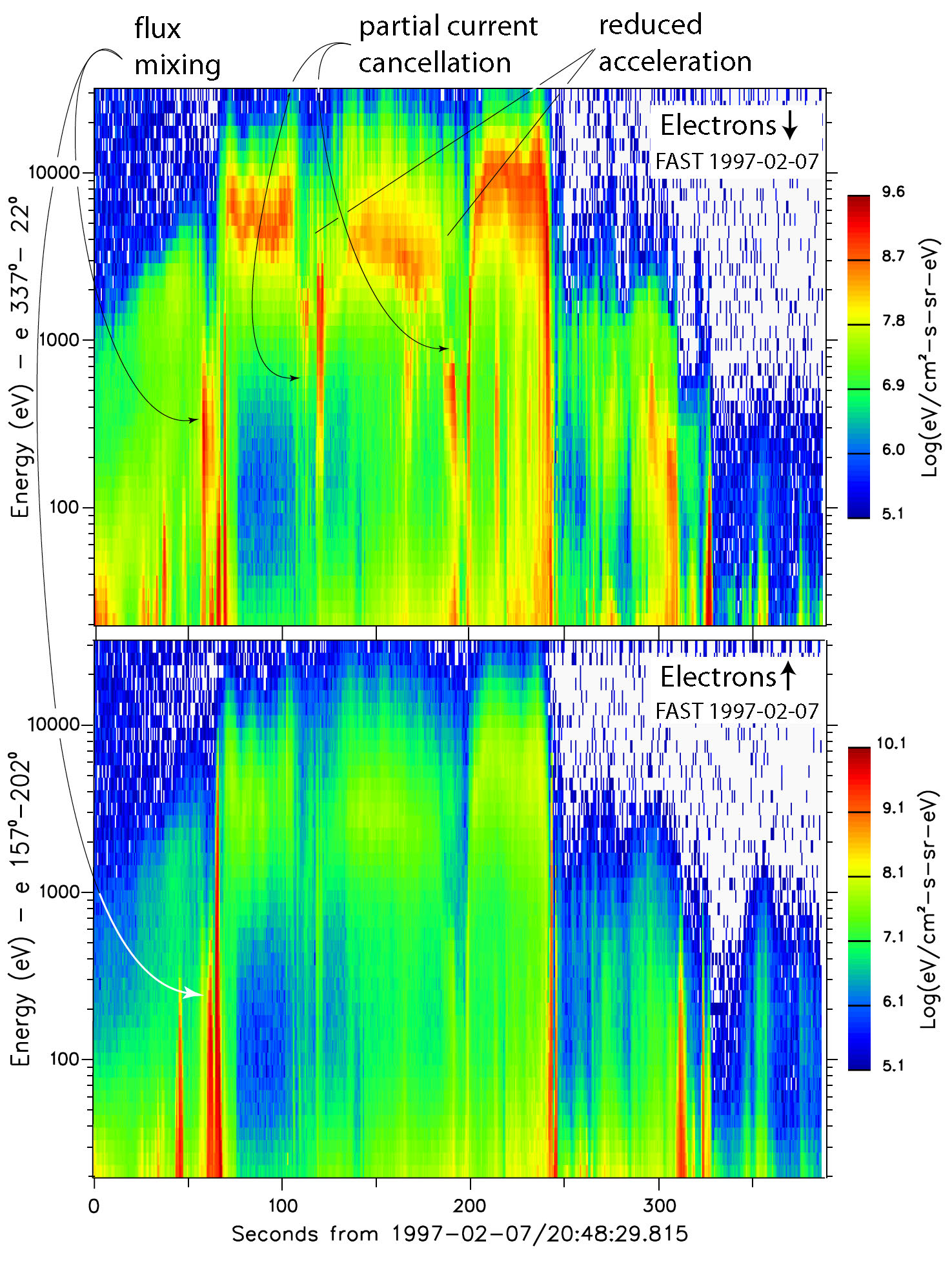}}
\caption{Sequence of downward (top)  and upward (bottom)  auroral electron fluxes observed by FAST on 02-07-1997 in the topside auroral region when crossing a substorm aurora \citep[after][]{treumann2011a}. The sequence distinguishes nicely between the intense  downward electron fluxes at energies $\epsilon_e\sim 10$ keV and downward fluxes at energies $\epsilon_e<0.1$ keV. Upward fluxes are confined to narrow spatial regions, downward fluxes are distributed over a much wider domain. In the transition regions between both domains one observes flux mixing which indicates that the current systems are not simply two dimension and also that there are many overlapping flux and current sources which the one-domensional path of the spacecraft does not resolve spatially. } \label{fig3}
\end{figure*}

\section{Reconnection under auroral conditions}
\begin{figure*}[t!]
\centerline{\includegraphics[width=0.75\textwidth,clip=]{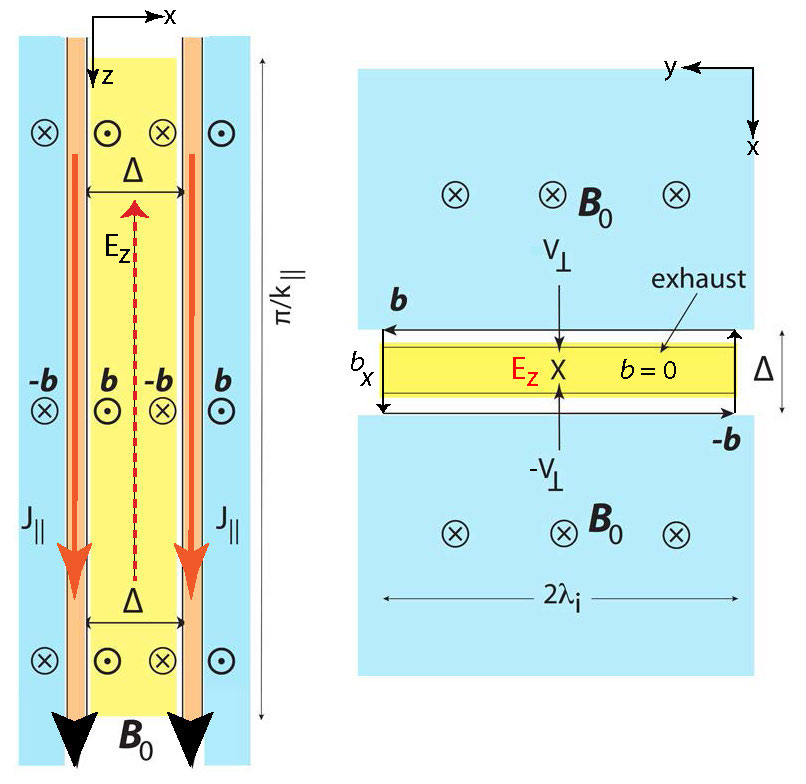}}
\caption{Schematic of the field configuration between two parallel field aligned flux tubes in the downward current region. \emph{Left}: Geometry along the ambient field. {Currents are in red. Included would be the (red dashed) fictitious return current which locally would correspond to the antiparallel wave magnetic fields $\pm b$. This current would be local over the wavelength of the inertial Alfv\'en wave. In any stationary reconnecting current picture it would be this current whose magnetic field reconnects. However, here this current does not exist in the exhaust. It is completely reconnected and gives rise to the reconnection electric field $E_z$ instead (dashed red) in the exhaust along the main magnetic field.  Electrons are directly accelerated by it along $\mathbf{B}_0$.} \emph{Right}: Reconnection geometry with perpendicular velocity $\mathbf{V}_\perp$, field free exhaust, reconnection fields $\pm b_x$ indicated, and $E_z$. {The anomalous collisions caused in the exhaust volume also permit for weak diffusion of the ambient field. This may cause what is believed to be magnetic field diffusion, a  very slow process compared to the wave/current induced spontaneous reconnection.}} \label{fig4}
\end{figure*}

Assuming stationarity of the field aligned current $\mathbf{J}=J_\|\mathbf{B}_0/B_0$ in two adjacent but separated parallel sheets and assuming, for simplicity, that the two currents are of equal strength, reconnection will  occur in the central region of separation of the current sheets. (Figure \ref{fig4} shows a two-cross section schematic of the downward current-field configuration for two closely spaced current sheets.)  Here the two magnetic fields of the field-aligned currents are antiparallel. According to the above discussion, we are in the downward current region \citep[as in the case of our previous radiation model][]{treumann2011}. In fact an analogue model would apply to the upward current region. The current flows in direction $z$; the direction of $y$ is longitudinal (eastward), $x$ is latitudinal (northward). If the sheet ist extended mostly in $y$ the antiparallel magnetic fields are in $y$ along the sheets. They will touch each other and reconnect between the two sheets thereby forming reconnection X-points with field component $\pm b_x$ and extended magnetic field free electron exhausts in $x$ and in $y$, which contain the local main-field parallel reconnection electric field, and accelerate electrons along the ambient field. Tjhese exhausts will propagate along the main field together with the wave. Plasmoids might also form in the separation between the sheets perpendicular to the ambient field,  and the presence of the strong ambient field will impose electron gyration and scatter of electrons  causing secondary effects like bursts of field aligned energetic electrons. Moreover, the exhausts will serve as source of radiation and various kinds of electrostatic instabilities (for instance Bernstein modes). 

{There are two essential differences between this type of reconnection and ordinary reconnection models. The first is that the ambient field serves as a strong guide field which, as noted, inhibits the adjacent field-aligned current to merge. The second is that initially the set-up lacks the presence of any central current sheet which in conventional models of reconnection is crucial and imposed from the beginning. In topside reconnection such a current flowing along the magnetic field inside the separation region would imply a return current which, however, is absent. Return currents flow through the bottomside ionosphere and close in the upward current region. Nevertheless formally a fictitious return current forms locally and temporarily in the centre of the separator, which can be assumed as distributed over the separating region and belonging to the antiparallel fields $\pm b$.}

{This current builds up dynamically and locally during the reconnection process itself when the two kinetic Alfv\'en waves slowly move perpendicular. This is a difficult dynamical problem in that reconnection will set on when the encountering magnetic fields exceed some threshold. Since electrons in this region are magnetized by the strong ambient field, they gyrate but do not take notice of the weak field $b$ of the kinetic Alfv\'en waves which is transported across the separating region by the perpendicular phase and group speeds of the waves to get into contact and merge.} 

{The reconnection process is thus solely between the two waves, primarily not affecting the ambient field and not based on any real central primary current sheet. Observations so far do not resolve the magnetic nor the particle effects of such fictitious return currents though some of the structure seen in the low energy electron fluxes in Fig. \ref{fig3} could be interpreted as such without proof. In fact, in order to avoid formation of the fictitious return current, which would imply that this current would be equally strong in the gap between the current sheets, reconnection is required over the full length of half a wavelength along $z$. Thus it necessarily generates elongated field-aligned vertical X-lines and electron exhausts in $z$.}

 \subsection{First step}
All these effects are of vital interest. However, one particularly interesting question concerns the dissipation produced by this kind of reconnection. It is frequently argued that it leads to sliding of main-field field lines. In order to understand such a mechanism one needs to know the anomalous resistance caused by reconnection. In electrodynamic formulation, reconnection is conventionally dealing solely with the merging and energy transfer of fields. The microscopic mechanism of energy transfer is accounted for in the transport coefficients. Hence the appropriate way of inferring their value is referring to the electromagnetic energy exchange. This leads to the application of Poynting's theorem
\begin{equation}
\frac{\partial b^2}{\partial t}=-\mu_0\,\eta_{an}J_\|^2-\nabla_\perp\cdot\big(\mathbf{E}_\|\times\mathbf{b}\big)-\nabla_\perp\cdot\big(\mathbf{E}_{rec}\times\mathbf{B}_0\big)
\end{equation}
where the contribution of the electric field to the left-hand side is neglected as it is relativistically small, and $\mathbf{b}$ is the magnetic field of the field-aligned current. It allows for a convenient estimate of the anomalous resistivity $\eta_{an}$ in reconnection without going into any microscopic detail of the mechanism of its generation. The electric field in this expression is along the ambient magnetic field, essentially being the electric field of the kinetic Alfv\'en wave. Estimates of this parallel field have been provided by \citet{lysak1996} and were taken as the important agent for accelerating auroral electrons. 

The above expression shows that reconnection in this case is a two-step process. In the first step the parallel field $E_\|$ along the ambient magnetic guide field sets up reconnection. In the second step the reconnection electric field $E_{rec}$ and exhaust have evolved. The cross-product with the main magnetic field then modifies the dynamics of the exhaust. 

\subsection{Anomalous collision frequency}
In this subsection we are not interested in this effect here as it is overwritten once reconnection really sets on but enters in the determination of the perpendicular inflow speed.  It causes it to be different from tailward reconnection. Instead we proceed to an estimate of the anomalous collision frequency.  

The parallel electric field $E_\|$ of the kinetic Alfv\'en wave plays an important role in the first step of the topside reconnection process. Since this field is parallel to the ambient geomagnetic field $\mathbf{B}_0$, the cross product with the wave magnetic field is responsible for the two current-sheet magnetic field components $\pm\mathbf{b}$ to approach each other in the region between the sheets. Hence, referring to this fact, the second term on the right can be expressed through the perpendicular velocity $\mathbf{V}_\perp=\mathbf{E}_\|\times\mathbf{b}/b^2$, and we have
\begin{equation}
\nabla_\perp\cdot\big(\mathbf{E}_\|\times\mathbf{b}\big)=\nabla_\perp\cdot\big(\mathbf{V}_\perp b^2\big)
\end{equation}

In order to get some information about the perpendicular velocity $\mathbf{V}_\perp$ which according to our coordinate system points to the centre of the region which separates the two current sheets, i.e. along $y$, we refer to the wave picture, noting that these pictures are equivalent: the field-aligned current $J_\|$ is carried by (upward topside ionospheric) electrons, on the other hand these electrons are transported (or pushed) by the kinetic Alfv\'en wave. {In fact, of course, $V_\perp$ is counted from each of the two parallel currents as pointing to the center of the separating sheet. It thus in our water-bag model changes abruptly sign in the center where due to the two antiparallel magnetic fields which collide there a fictitious weak return current of strength $j_\|\approx 2b/\mu_0\delta$ arises, with $\delta$ the fictitious width of this narrow current layer which we do not explicitly consider. The simplest is in our water-bag model to assume that for closely separated parallel current sheets we have essentially 
$\delta\to\alpha\Delta$, with $\alpha\lesssim1$, and a return current distributed over almost the entire separation width. One should also keep in mind that any field-aligned current carried by the Alfv\'en wave is a current pulse with both $E_\|$ and $b$ changing direction (oscillating) over half the wavelength. Thus $V_\perp$ for each current pulse on one ambient field line has same sign over the full wave length while maximizing twice. On using this equivalence the perpendicular velocities $\pm V_\perp$ are just the perpendicular phase speeds of the kinetic Alfv\'en waves on the two adjacent current sheets}
\begin{equation}
V_\perp\sim\frac{\omega}{k_\perp}\approx \frac{V_A}{\sqrt{1+k_\perp^2\lambda_e^2}}\frac{k_\|}{k_\perp}\ll V_A
\end{equation}
{This velocity apparently diverges for $k_\perp\to0$ which, however, is not the case because the kAW is a surface wave being defined only for $k_\perp\neq0$ while becoming a body wave for $k_\perp\to0$ carrying no current anymore. Its most probable wavenumber is about $k_\perp\lambda_e\sim1$ attributing to the parallel phase and group velocity $\sim V_A/\sqrt{2}$ and a perpendicular group velocity $\sim-V_A(k_\|\lambda_e)/2^{3/2}$. However, since $V_\perp$ indeed transports not only the field but also energy, one may argue that the use of the latter expression would be more appropriate than the phase speed. Since this does not make any big difference for our purposes, we in the following for reasons of simplicity understand $V_\perp$ as phase speed. For more precise expressions one may replace it in the following with the perpendicular group speed}. 

The velocity $V_\perp$ is small because $k_\|\ll k_\perp$, i.e. the kinetic Alfv\'en wave is long-wavelength parallel to the ambient field but of short perpendicular wavelength, a very well-known property. Moreover, $V_\perp(z)$ may vary along the ambient field but, in the frame of the wave, which corresponds to a water-bag model, is constant in the perpendicular direction. Hence, of the above vector product just remains the variation of the magnetic field $b(x)$ over the distance between the two current sheets. This insight enables us to rewrite Poynting's equation as
\begin{equation}
\frac{\partial b^2}{\partial t}\approx -\mu_0\,\eta_{an}J_\|^2-V_\perp\frac{\partial b^2}{\partial x}
\end{equation}
which, assuming a stationary state, enables us to estimate the anomalous resistivity of stationary reconnection (in the wave frame) where the inflow of magnetic energy attributed by the current, i.e. the field-aligned electron flux whose origin is found in reconnection in the magnetotail, is balanced by anomalous energy transfer to the plasma in the region separating the two current sheets. Putting the left-hand side to zero we thus find that in this kind of topside reconnection the anomalous resistivity is bound from above as
\begin{equation}
\eta_{an}\lesssim \frac{4V_A}{\sqrt{1+k_\perp^2\lambda_e^2}}\frac{k_\|}{k_\perp}\frac{b^2}{\mu_0 \Delta J_\|^2}{\approx \frac{2V_Ab^2}{\mu_0\Delta J_\|^2} \frac{k_\|^2\lambda_e^2}{(1+k_\perp^2\lambda_e^2)^{3/2}}}
\end{equation}
where $\Delta$ is the spatial separation of the two field-aligned current sheets, and we have taken into account that each of the two identical current layers contributes a field $b$. {The second part of this expression makes use of the perpendicular group speed. This resistivity is small as $k_\|^2\lambda_e^2/\Delta$ but finite. It gives rise to a finite diffusion coefficient that can be interpreted as an anomalous diffusivity for the ambient magnetic field in the auroral topside ionosphere, caused by topside reconnection between anti-parallel current sheets in the downward current region. We might note at this occasion that the restriction to the downward current region is motivated by the observation of narrow current sheets in the downward current region. Observations do not suggest that similarly narrow current sheets evolve in the upward current region. If this would be the case, the same arguments would apply there, causing reconnection and a similar anomalous resistivity.}  

What concerns the spatial separation of the current sheets (see Fig. \ref{fig3}), the best available observations (FAST) do not resolve any single sheets; it can however be assumed that their scales are the order of or below the ion-inertial length, such that $\Delta\lesssim \lambda_i\sim$ several to many $\lambda_e$. This may overestimate the real value but has been accounted for in writing the expression as an upper limit. Determination of the anomalous resistivity thus requires knowledge of the field aligned current density, current sheet separation, and the transverse magnetic field component of the sheet current. We then can estimate the anomalous collision frequency $\nu_{an}= \eta_{an}\epsilon_0\omega_e^2$ in this kind of reconnection
\begin{equation}
\nu_{an}\lesssim \frac{V_A/\alpha^2\lambda_e}{\sqrt{1+k_\perp^2\lambda_e^2}}\frac{k_\|\Delta}{k_\perp\lambda_e }
\end{equation}
where we used that $J_\|\approx 2b/\alpha\Delta$. Note that $V_A=B_0/\sqrt{\mu_0 m_i N}$ is based on the ambient magnetic field and plasma density. This simple estimate shows that reconnection in this case can, under stationary condition be described as being equivalent to a diffusive process based on the anomalous collision frequency which is provided by the merging of the transverse magnetic fields of the two neighbouring field-aligned current sheets. Since the related diffusivity is felt in the entire region it is remarkable that it could effect also the main ambient guide field. In other words, topside reconnection could become responsible for diffusion of the main magnetic field lines in a locally restricted domain possibly causing effects on a larger scale in the auroral region.   

Real reconnection will not occur between field-aligned current sheets of same strength. Thus the above resistivity respectively the collision frequency must be reduced by another factor proportional to the involved current and field fractions. 

\subsection{Second step: Reconnection electric field}
{So far we just investigated the energy balance in order to obtain an anomalous collision frequency in this kind of reconnection. Reconnection however manifests itself in X points generating transverse magnetic fields and in addition electric fields. Since there is no primary return current flowing, it cannot be used as input into the two-dimensional reconnection equation for the vector potential $A_z$
\begin{equation}
\nabla^2 A_z=-\mu_0j_z(x), \qquad \nabla= (\partial_x,\partial_y,0), \qquad j_z(x)= -2\epsilon b(x)/\mu_0\Delta
\end{equation}
without prescribing the built-up of the central current profile $j_z(x)$, which is possible only when assuming that the $b$ is independent of $x$, in which case it provides the usual stationary tearing mode solution \citep[see, e.g.,][]{schindler1974} rewritten for electrons alone. Under these simplifying restrictions the two components of the reconnected magnetic field including the X point are given by $\mathbf{b}=(\partial_yA_z,-\partial_xA_z,0)$, which to refer to suffices for our qualitative considerations. The a priori assumption of a return current is, however, incorrect. On the topside there may weak local return currents exist filling the separations between the narrow downward current sheets, but the main return current flows in the upward current region and is distributed over a wide domain. Hence just a fraction $\epsilon$ of return current can flow in the gap, as included in the last expression. The electric field in this case primarily has only one component, which is along the main field and is given by $E_z=-\partial_t A_z-\nabla U$ where $U$ is the scalar electrostatic gauge potential which may occur if an inhomogeneity exists or the system is not ideally symmetric. This field adds to the field aligned kinetic Alfv\'en wave electric field and contributes to electron acceleration. It is the wanted reconnection electric field and can be much larger than the small linear wave electric field. Unfortunately its precise knowledge  requires solution of the equation for the vector potential $A_z$ and some interpretation of the time derivative operator. The latter can be transformed into a spatial derivative $\partial_t= \pm\mathbf{V}_\perp\cdot\nabla$, still requiring the solution $A_z(x,y)$. }

{The important conclusion in the case of topside reconnection is rather different from usual reconnection. It tells that the exhaust is, over half the wavelength of the inertial Alfv\'en wave free of wave magnetic fields $b$, while being bounded by the reconnected wave fields $\pm b_x$. The exhaust instead contains the reconnection electric field, by being along the main field, does directly contribute to acceleration respectively deceleration of electrons  (and also ions) along the main magnetic field, one of the most important and still unresolved problems in auroral physics. There acceleration is attributed to a variety of waves, reaching from kinetic Alfv\'en through whistlers and several electrostatic waves to electron and ion holes. Except for the latter nonlinear structures, all wave electric fields are quite weak, and in addition fluctuate. Acceleration thus becomes a second order process.} 

{In case of the topside reconnection, a mesoscale first-order electric field $E_z$ is produced which directly accelerates particles, depending on its direction along the main field. Moreover, the source of the accelerated particles is the gap region between the two current sheets, the so-called exhaust, such that the kinetic Alfv\'en wave electric field and the reconnection electric field do barely interfere. Hence the full strength of the reconnection exhaust field acts accelerating. One may thus conclude that topside reconnection, if it takes place, will substantially contribute to auroral particle acceleration. 
 }

{In order to circumvent the above named difficulty of calculating $A_z$ and to obtain an estimate of the reconnection electric field, we may return to the induction equation in its integral form where the electric field is given by the integral over the surface of the reconnection site
\begin{equation}
\oint \mathbf{E}\cdot d\mathbf{s}=-\frac{d\Phi}{dt}= -\frac{d}{dt}\int\mathbf{b}\cdot d\mathbf{F}
\end{equation}
and the right-hand side is the exchange of magnetic flux in the reconnection process within the typical time $dt=\tau_{rec}$. This time is not necessarily the same as the anomalous collision time. The magnetic flux is given by 
$\Delta\Phi\approx 4\pi b\Delta/k_\|$. The line integral over the boundary of the reconnection site becomes $\approx4\pi E_z/k_\|+2\Delta \delta E_x$. Under ideally symmetric conditions the second term would vanish because the two contributions of the $x$ integration would cancel out. If some asymmetry is retained then a finite component $\delta E_x$ arises. Taken these together yields dimensionally (not caring for the signs)
\begin{equation}
4\pi E_z/k_\|+2\Delta \delta E_x \approx 4\pi b\Delta/k_\|\tau_{rec}
\end{equation}
Neglecting the small second term on the left then gives a simple order of magnitude estimate of the reconnection electric field
\begin{equation}
E_z\approx \frac{b\Delta}{\tau_{rec}}
\end{equation}
which could have been guessed from the beginning. This contains the reconnection time $\tau_{rec}$ which  so far is undetermined. It can be taken for instance as the above derived anomalous collision time $\tau_{an}=\nu_{an}^{-1}$. Below we derive another characteristic time. Which one has to be chosen, cannot decided from these theoretical order of magnitude  estimates. It is either due to observation or numerical simulations.} 

The small additional term $2\Delta\delta E_x =-U$ is a potential field produced by a possibly present asymmetry between the original current sheets or some gradient in the particle density. Such a gradient can be produced, if a substantial part of the electron component in the gap is accelerated away along the main field, causing a dilution of plasma in the exhaust. Being perpendicular to the magnetic fields $\mathbf{B}_0$ and $\mathbf{b}$ it leads to weak shear motions and circulation of the electrons inside the gap-exhaust region, which should observationally be detectable.

\subsection{Reconnection time}
In the above we have made use of the notion of reconnection time $\tau_{rec}$. Here we attempt a clarification of this time. Topside reconnection will not be stationary. It should vary on the time scale of the kinetic Alfv\'en frequency respectively moving together with the latter along the magnetic field. This motion should mainly be upward since causality requires that the wave transports information back upward with the upward moving electrons in the downward current region. It will thus be modulated and lead to quasi-periodic acceleration and generate medium energy electron bursts ejected from the local electron exhaust reconnection region along the sheet current magnetic field. These bursts flow perpendicular to the ambient field, start gyrating and immediately become scattered along the ambient field spiralling mainly upward into the weak ambient field region. Their pitch-angle distribution should obey a well defined downward loss-cone. 

With the above estimate of the anomalous resistivity in this kind of reconnection, we can proceed asking for the typical reconnection time scale. For this purpose we return to Poynting's full theorem and take its variation with respect to the stationary state, indexing the latter with 0 while keeping the slow perpendicular velocity $V_\perp$ fixed but varying the resistivity. We need to express the parallel current through the resistivity. This can be done via the electric field $E_\|$ to obtain
\begin{equation}
J_\|^2=\eta^{-2}E_\|^2=\eta^{-2}b^2V_\perp^2
\end{equation}
This procedure, after some straightforward and simple algebra and rearranging, leads to the following expressions
\begin{eqnarray}
\frac{d(\delta b)^2}{dt}\equiv\Big(\frac{\partial}{\partial t}+V_\perp\nabla_\perp\Big)(\delta b)^2&=&- 2\mu_0J_{\|0}^2\delta\eta\\
\delta\eta&=&-\frac{V_\perp}{\mu_0J^2_{\|0}\Delta} (\delta b)^2
\end{eqnarray}
and we obtain dimensionally for the typical time of reconnection
\begin{equation}
\tau_{rec}\sim\frac{2\Delta}{V_\perp}
\end{equation}
This seems a trivial result, but it tells that reconnection is a process which annihilates the excess magnetic field which is provided by the perpendicular inflow under the condition that we are close to a stationary state. This time can be compared with the times of energy flow in the shear Alfv\'en wave. Since clearly $V_\perp\approx\partial\omega/\partial k_\perp$, one obtains
\begin{equation}
\tau_{rec}\approx\frac{2\Delta}{V_A}\frac{k_\perp}{k_\|}\frac{\big(1+k_\perp^2\lambda_e^2\big)^\frac{3}{2}}{k_\perp^2\lambda_e^2}
\end{equation}
a time the length of which depends essentially on the spacing of the current sheets. Since $V_A$ is large, there will be a balance between the spacing and the domain of the kinetic Alfv\'en wave spectrum which allows reconnection to occur in the topside. Let the vertical topside width be $L_z$ and the Alfv\'en time $\tau_A=L_z/V_A$ then we have the condition 
\begin{equation}
\frac{\tau_{rec}}{\tau_A}\approx \frac{2\Delta}{L_z}\frac{k_\perp}{k_\|}\frac{\big(1+k_\perp^2\lambda_e^2\big)^\frac{3}{2}}{k_\perp^2\lambda_e^2}< 1
\end{equation}
for reconnection to occur in topside parallel field-aligned current sheets. This essentially is a condition on the spacing $\Delta$ of the sheets, meaning that 
\begin{equation}
\frac{\sqrt{32}\,\Delta}{L_z}< \frac{k_\|}{k_\perp}=\frac{\lambda_\perp}{\lambda_\|}\ll 1
\end{equation}
Any current sheet separation is strictly limited. Since it must be larger than the upward electron gyro-radius we have $\Delta>r_{ce}$. Both conditions are easily satisfied.

\subsection{Conclusions}
{In the present letter we propose that reconnection might occur not only in given current sheets but also in the topside ionosphere-magnetosphere auroral transition region  where the main magnetic field is very strong, almost vertical, and directly connects to the tail reconnection region. It serves  as a guide for any particle flow exchange between the topside ionosphere and the tail plasma sheet, exchange between low frequency electromagnetic waves (in our case kinetic Alfv\'en waves) trapped in flux tubes and the accompanying field-aligned current sheets, and ultimately as an inhibitor for the field-aligned parallel current sheets to merge. This  enables reconnection in the gap between the current sheets between the oppositely directed magnetic field of the sheets respectively the kinetic Alfv\'en wave magnetic fields.} 

{Dealing with reconnection, one is not primarily interested in the change of magnetic topology but in energy transformation from magnetic into kinetic, diffusion of plasma and magnetic field across the reconnection region, generation of electric fields, and ultimately selective particle acceleration as these are the observed effects. }
The generality of reconnection is not the best argument. The decades old claim that reconnection converts magnetic energy into mechanical energy is no fundamental insight; in all processes involving reconnection, the main energy is stored in the basic mechanical motion and by no means in the magnetic field. This motion, convection in inhomogeneous media with boundaries, like the magnetotail or the magnetopause, or turbulence necessarily produces currents and transports magnetic fields to let them get into contact. The amount of energy released by reconnection is in all cases just the minor electromagnetic part, a fraction of the mechanical energy. 

{Topside reconnection is expected predominantly in the downward current region, which observationally seems to be highly structured, consisting of several adjacent parallel current sheets. Similar  conditions may also occur in the upward current region though no such structuring is obvious from observations. If it exist, then the physics will be similar. We have shown that topside reconnection is possible, generates a elongated field-aligned regions (exhausts) where the fields of parallel current sheets merge, anomalous collisions are generated, energy is exchanged and dissipated, and most important a first order reconnection electric field $E_{rec}$ is produced in the exhaust along the ambient magnetic field but  restricted to the gap region between the current sheets. This field is capable of accelerating electrons along the main field, as is most desired by all auroral physics. Here it comes out as a natural result of topside reconnection.} Topside reconnection generates parallel electron beams, it lifts the escaping electrons in the exhaust into an elevated parallel energy level. These beams then cause a wealth of auroral effects in the environment and when impinging onto the upper ionosphere. Acceleration of electrons by the reconnection electric field leaves behind an electron depleted exhaust mainly containing only an anisotropic electron component whose pitch angle distribution peaks at perpendicular energies.  

{It is instructive to briefly inspect Fig. \ref{fig3}. It shows the downward (upper panel) and upward (lower panel) electron fluxes. In addition to the temporally/spatially highly structured fluxes, still obeying the spatial differences between the downward and upward current regions imposed by the tail-source of the downward fluxes, resulting  from  variations in tail-reconnection, or several tail-reconnection sites, one occasionally observes the simultaneous presence of upward and downward fluxes in the downward current region. One particular case it at $t\approx 60$ s. The upward electron fluxes maximize below $\sim0.1$ keV. Simultaneously a banded flux of downward electrons with central energy $\sim 0.3$ keV appears in the upper panel. This event  is indicated as flux mixing. It could also be understood as acceleration of electrons resulting from the local reconnection in the gap between current sheets. }

Aside of acceleration, radiation generation may be taken as signature of topside reconnection. Radiation is preferrably generated by the electron cyclotron maser mechanism. It requires low electron densities, strong magnetic fields, and a rather particular particle distribution with excess energy in its component perpendicular to the ambient magnetic field \citep{sprangle1977,melrose1985}. Such a state in dilute plasmas lacks sufficiently many electrons for re-absorbing the spontaneously emitted radiation while the excited state causes inversion of the absorption coefficient. These conditions allow for the plasma to become an emitter \citep{twiss1958,schneider1959,gaponov1959} by the electron cyclotron maser mechanism \citep{wu1979} based on a loss-cone distribution \citep{louarn1996}. It requires weakly relativistic electrons \citep[see][for reviews]{melrose1985,treumann2006} and a low density electron background embedded into a strong field. It nicely comes up for the weak auroral kilometric background radiation but fail explaining the intense narrow band observed and drifting emission seen in panel $d$ of Fig. \ref{fig2}. 

To explain the latter, in earlier work we referred to electron hole formation \citep{pottelette2005,treumann2011}. Hole models favourably apply to electron depleted exhausts in topside reconnection where densities become low \citep[see, e.g.][]{treumann2013} and the remaining trapped electron component  maximizes at perpendicularly speeds having large anisotropy. Intense narrow band drifting emissions in the frequency range 300-600 kHz may be a signature of topside reconnection in the strong main auroral field. They were originally attributed to Debye scale electrostatic electron holes \citep{ergun1998b,pottelette1999} observed by Viking \citep{deferaudy1987} and  FAST \citep{carlson1998,ergun1998a,pottelette2005} but are to small-scale for radiation sources. Topside reconnection exhausts instead have dimensions along the magnetic field of half a kinetic Alfv\'en length and transverse scales of few ion inertial lengths $\lambda_i$ or $\sim 100\lambda_e$. Such scales can host and amplify one or more radiation wave lengths.

Of course, details of this process should be developed both analytically as far as possible, and by numerical simulations. If confirmed, this mechanism would also map to any astrophysical moderately or strongly magnetized object with appropriate modification. 

The present qualitative considerations which we spiced with a few simple estimates based on energy conservation arguments just propose that reconnection in the topside auroral ionosphere is a process which has so far been missed and probably is that mechanism which releases the largest amount of so-called magnetically stored energy available and from the smallest spatial regions. Reconnection in much weaker fields like in turbulence and broad current sheets will be substantially less efficient because of the weakness of the reconnecting magnetic fields. Nevertheless in very large extended systems with reconnection proceeding on the microscales  \citep{treumann2015} with the total number of reconnection regions very large, the emission measure is large as well, and radiation from reconnection may become a non-negligible signature even in weak fields. However, in very strong fields like those in magnetized planets and magnetized stars (predominantly neutron stars, white dwarfs but also including outer atmospheres of magnetized stars like the sun) reconnection following our argumentation may be more important than so far assumed.

\section*{Acknowledgments}
 This work was part of a brief Visiting Scientist Programme at the International Space Science Institute Bern. RT acknowledges the interest of the ISSI directorate as well as the generous hospitality of the ISSI staff, in particular the assistance of the librarians Andrea Fischer and Irmela Schweitzer, and the Systems Administrator Saliba F. Saliba. We acknowledge discussions with R. Nakamura, and Y. Narita. RT acknowledges the cooperation with R. Pottelette two decades ago  on the data reduction and the radiation and electron hole problems.

\end{document}